\documentclass[journal]{IEEEtran}

\usepackage{cite}
\usepackage{setspace}
\usepackage{graphicx}
\usepackage{array}
\usepackage{float}
\usepackage{tikz}
\usepackage{ifthen}
\usetikzlibrary{shapes.geometric}
\usetikzlibrary{decorations.pathreplacing}
\usetikzlibrary{positioning, calc}
\usetikzlibrary{patterns}
\usepackage{pgfplots}
\usepgfplotslibrary{fillbetween}
\usepackage{pstool}  
\usepackage{mathtools}
\usepackage[linesnumbered,algoruled,boxed,lined]{algorithm2e}
\usepackage[absolute]{textpos}
\usepackage{threeparttable}
\usepackage[strict]{changepage}
\usepackage{flushend} 
\usepackage{amssymb}
\usepackage{comment}
\usepackage{bbm}
\usepackage{hhline}
\usepackage{lipsum} 
\usepackage{enumitem}
\usepackage{soul}
\usepackage{xcolor}

\usepackage{amsthm}

\newcommand{\review}[1]{{\color{black}#1}}

\ifCLASSINFOpdf
\else
\fi
\usepackage{amsmath}
\ifCLASSOPTIONcompsoc
  \usepackage[caption=false,font=normalsize,labelfont=sf,textfont=sf]{subfig}
\else
  \usepackage[caption=false,font=footnotesize]{subfig}
\fi
\pgfplotsset{compat=1.14} 

\begin{document}
%
\title{Adaptive Compression of Massive MIMO Channel State Information with Deep Learning}
%

\author{Faris~B.~Mismar,~\IEEEmembership{Senior~Member,~IEEE}
    ~and Aliye \"{O}zge Kaya,~\IEEEmembership{Senior~Member,~IEEE}
\thanks{The authors are with Nokia Bell Labs Consulting and Nokia Bell Labs. Email: faris.mismar@bell-labs-consulting.com and ozge.kaya@nokia-bell-labs.com.}
}%

\maketitle
\begin{abstract}
This paper proposes the use of deep autoencoders to compress the channel information in a \review{massive} multiple input and multiple output (MIMO) system.  Although autoencoders perform lossy compression, they still have adequate usefulness when applied to \review{massive} MIMO system channel state information (CSI) compression.  To demonstrate their impact on the CSI, we measure the performance of the system under two different channel models for different compression ratios. \review{We disclose a few practical considerations in using autoencoders for this propose}.  We show through simulation that the run-time complexity of this deep autoencoder is irrelative to the compression ratio \review{and thus an adaptive compression rate is feasible} with an \review{optimal} compression ratio depending on the channel model and the signal to noise ratio.  
\end{abstract}


%
\IEEEpeerreviewmaketitle

\section{Introduction}\label{sec:introduction} 
%
%
%
%



 
One of the main areas that the sixth generation of wireless communications (6G) focuses on is an artificial intelligence (AI)-native air interface.  As users are stationary or move at low speeds, the channel state information (CSI) changes slowly due to larger channel coherence times.  Further, since different orthogonal frequency division multiplexing (OFDM) subcarriers experience correlated fading despite the different frequencies \cite{bandswitching}, applying parts of the CSI towards other OFDM subcarriers and removing \textit{redundant} others (or ``compressing'' the channel) is beneficial. Also, with neural receivers being a quintessential technology for 6G, it makes sense to employ these neural networks for compression \cite{3gpp38843}\review{---especially in a massive multiple-input multiple-output (MIMO) setup.}


When it comes to the use of AI towards air interface radio resource management, there is no shortage of literature\review{---even in the CSI compression use case}.  For example, the use of long short-term memory deep learning models were used to predict the next power-optimal beam for users on a trajectory was studied in \cite{10122528}.  In \cite{bandswitching, 8938771}, the successful use of deep neural networks (DNNs) in a supervised learning mode to improve handovers across different frequency bands and introduce power control was demonstrated.  Furthermore, the use of supervised learning to perform CSI compression in MIMO systems was researched in \cite{9466243}.  A comprehensive overview of CSI compression and reconstruction using deep neural networks was studied in \cite{9931713, 9691478}; however, insights on usability in light of performance and compression ratio were lacking.  An attempt to estimate the CSI using a DNN was studied in \cite{10230266} but without the compression effect.  Further \cite{8802261} studied the impact of compression on multiple input and single output channel configurations using non-DNN related compression methods.  \review{Furthermore, the dominant measure of compression performance in prior work focuses on the normalized mean square error (in dB) as the \textit{de facto} measure of system performance and thus compression ratios, whereas our focus is on a set of practical measures of error that capture the experience of the payload and are used in realistic deployments of cellular networks.}

\begin{figure}[!t]
\centering
\resizebox{0.45\textwidth}{!}{\begin{tikzpicture}[style=thick, node distance=2cm, scale=2, >=latex]
    \node [rectangle, draw, dashed,
            text width=4.5em, text centered, minimum height=2em] (estimator) {Estimation}; 
    \node [coordinate, above=of estimator.40, yshift=-3em] (pilot_out) {};
    \draw [draw,->] (pilot_out) -- node [right] {$\mathbf{Y}_\text{pilot}$} (estimator.40);
    \node [coordinate, above=of estimator.140, yshift=-3em] (pilot_in) {};
    \draw [draw,->] (pilot_in) -- node [left] {$\mathbf{X}_\text{pilot}$} (estimator.140);

    \node [rectangle, draw, line width=0.75mm, right of=estimator, fill=white!85!black, node distance=2.5cm,
            text width=3em, text centered, minimum height=2em] (compchannel) {$\mathsf{C}_\kappa(\cdot)$};
    
    \draw [draw,->] (estimator) -- node[above] {$\mathbf{\hat H}$} (compchannel);

     \node [rectangle, draw, line width=0.75mm, right of=compchannel, fill=white!85!black, node distance=2cm,
            text width=3em, text centered, minimum height=2em] (quantizer) {$Q(\cdot)$};

      \node [rectangle, draw, dashed, right of=quantizer, fill=white!85!black, node distance=2cm,
            text width=3em, text centered, minimum height=2em] (decompchannel) {$\mathsf{C}^{-1}_\kappa(\cdot)$};
    
    \node [rectangle, draw, line width=0.75mm, right of=decompchannel,
            text width=3em, text centered, minimum height=2em] (precoder) {$\mathbf{F}$}; 

    \node [rectangle, draw, dashed, below of=precoder, node distance=1cm,
            text width=3em, text centered, minimum height=2em] (combiner) {$\mathbf{G}$}; 
    
    \draw [draw,->] (compchannel) -- node {} (quantizer);
    \draw [draw,->] (quantizer) -- node {} (decompchannel);

\draw [draw,->] (decompchannel) -- node[above, xshift=0.5em] {$\mathbf{\hat{\hat H}}^{(\kappa)}$} (precoder);

   	\node [coordinate, fill, circle, inner sep=1pt, right of=decompchannel, xshift=-3em] (output2) {};

   \draw [draw,->] (output2) |- node {} (combiner);
    
\end{tikzpicture}%
}%
\caption{An overview of channel compression onto the estimated channel $\mathbf{\hat H}$ with channel compression $\mathsf{C}_\kappa$, quantization $Q$, and decompression $\mathsf{C}_\kappa^{-1}$ functions (in gray).  Precoder $\mathbf{F}$ and combiner $\mathbf{G}$ are computed from the reconstructed channel $\mathbf{\hat{\hat H}}$.  Dashed (thick) borders represent functions that are computed at the user (base station) side.}%
\label{fig:overall}
\end{figure}%

Due to the limited \review{practical} insights in the literature on the applicability of self-supervised learning CSI compression in \review{a massive} MIMO system using DNNs in terms of \review{bit and block error} performance, we have written this paper proposing a solution that compresses the CSI in a MIMO system.   Fig.~\ref{fig:overall} shows the proposed solution overall.  The main contributions of this paper can be summarized as:

\begin{enumerate}
\item Formulating the \review{6G-relevant massive} MIMO channel estimate compression problem as a deep learning problem.
\item Describing the channel conditions in which such a compression can be efficiently used and the compression ratio vs.\ performance trade-off.
\review{\item Implementing a variant with \textit{adaptive} compression with an optimal compression ratio computed based on CSI.}
\end{enumerate}%

The rest of this paper is structured as follows: Section~\ref{sec:sysmodel} describes the system model.  The application of autoencoders to perform channel compression is motivated in Section~\ref{sec:autoencoders}.  In Section~\ref{sec:performance}, several relevant performance measures are introduced.  Section~\ref{sec:simulation} shows results of the simulations performed.  Finally, Section~\ref{sec:conclusion} concludes the paper.

\vspace*{-0.5em}
\section{System Model}\label{sec:sysmodel}
We consider a \review{massive} MIMO-OFDM system model with $N_t$ transmit antennas at the base station (BS) and $N_r$ receive antennas at the user equipment (UE) and $N_\text{sc}$ OFDM subcarriers with spacing of $\Delta f$.  \review{The transmit antennas have a uniform rectangular array (URA) shape with half a wavelength element spacing}.  Let the transmit signal power be $P_x$. This \review{massive} MIMO system for the transmitted OFDM subcarriers is thus: 
\begin{equation}\label{eq:system}
    \mathbf{y} = \mathbf{H}\mathbf{x} + \mathbf{n}
\end{equation}%
where $\mathbf{y} \in \mathbb{C}^{N_r}$ is a column vector containing received OFDM subcarriers from a transmitted column vector $\mathbf{x} \in \mathbb{C}^{N_t}$ such that $\mathbb{E}[\Vert\mathbf{x} \Vert^2] \coloneqq P_x = E_x\Delta f$, $\mathbf{H}\in \mathbb{C}^{N_r\times N_t}$ is the true channel state information (CSI) matrix, and $\mathbf{n}\in \mathbb{C}^{N_r}$ is a column vector of additive noise the entries of which are independent and identically sampled from a zero-mean complex Normal distribution $\mathbf{n}\sim\mathcal{N}_\mathbb{C}(0, \sigma_n^2)$, with the noise power being the power of the OFDM subcarrier bandwidth and the noise power spectral density (i.e., $\sigma_n^2 = N_0\Delta f$). The number of streams $N_s \coloneqq \min(N_t, N_r)$.  Finally, we encode the transmitted vector $\mathbf{x}$ with cyclic redundancy check (CRC) for error detection.



\textbf{Fading:}  To construct a \textit{true} (vs.\ learned) fading channel $\mathbf{H}$, we employ industry standards clustered delay line (CDL) channels as defined in \cite{3gpp38900}.  We use the delay profiles \mbox{CDL-C} and \mbox{CDL-E} with Cluster 1 channel model, which are used for modeling non-line-of-sight (NLOS) and line-of-sight (LOS) propagation respectively.

\review{\textbf{Sparsification:} The aforementioned channels are sparse due to the limited number of multipath components in both LOS and LOS propagation.  Such limited number of multipath components contributes to the sparsity of the channel.}

\textbf{Channel estimation:}  The BS sends a sequence of pilots $\mathbf{X}_\text{pilot}\in\mathbb{C}^{N_\text{pilot}\times N_t}$ to the UE.  The UE estimates the channel based on these pilots sent by the BS and the received pilots using a least square channel estimator.  That is:
\begin{equation}\label{eq:estimator}
    \mathbf{\hat H} = \mathbf{Y}_\text{pilot}\mathbf{X}_\text{pilot}^\ast(\mathbf{X}_\text{pilot}\mathbf{X}_\text{pilot}^\ast)^{-1}.
\end{equation}%
For the purpose of differentiating the channels of multiple users, we use a subscripted variant $\mathbf{\hat H}_v$, with $v\in \{1,2,\ldots,N_\text{UE}\}$ for the $v$-th UE.

\textbf{Channel compression and decompression:}  The UE compresses the CSI and sends it to the BS.  The BS decompresses and reconstructs the CSI for subsequent transmission within the channel coherence time.  Denoting the compression operation with parameter $\kappa$ as $\mathsf{C}_\kappa$ and the quantizer as $Q(\cdot)$:
\begin{equation}\label{eq:reconstructed_csi}
\mathbf{\hat{\hat H}}^{(\kappa)} = \mathsf{C}_\kappa^{-1}(Q(\mathsf{C}_\kappa(\mathbf{\hat H}))),
\end{equation}%
where $\mathbf{\hat{\hat H}}^{(\kappa)}$ is the \textit{reconstructed} CSI after a compression with parameter $\kappa$.  More on compression in Section~\ref{sec:autoencoders}.

\textbf{Precoding and combining:} The BS performs a singular value decomposition on the reconstructed CSI estimate matrix after decompression.  This leads to the two matrices: 1) the precoding matrix $\mathbf{F}\in\mathbb{C}^{N_t\times N_s}$ which is computed at the BS and allows the setting the transmit power per channel eigenmode through waterfilling and 2) the combining matrix $\mathbf{G}\in\mathbb{C}^{N_t\times N_r}$ which is applied at the receiver (the UE).  This effectively diagonalizes the CSI into a matrix $\boldsymbol\Sigma\in\mathbb{C}^{N_s\times N_t}$.

\textbf{Users:} UEs are scattered in the service area of the BS and each has a number of an antennas equal to $N_r \ge 1$. These UEs move at pedestrian-like speeds within the BS service area.

\textbf{Channel equalization:}  The decompressed channel is equalized at the receiver.  The equalization matrix $\mathbf{W}\in \mathbb{C}^{N_t\times N_r}$ is left-multiplied by the received signal to obtain $\mathbf{z}$ which is passed through a maximum likelihood detector to recover the received symbols $\mathbf{\hat x}$.  Equalization also amplifies the noise $\mathbf{v}\coloneqq \mathbf{Wn}$ at the receiver.

\textbf{Statistics:} The reference symbol received power is the received power of a given OFDM resource element and is equal to $P_x/(N_tN_\text{sc})$.  Furthermore, the transmit OFDM subcarrier signal to noise ratio (SNR) is $\rho \coloneqq E_x/(N_\text{sc}N_tN_0)$. 
 All statistics are measured during the channel coherence time.  We show the channel compression loss, which is the autoencoder cost function as in Section~\ref{sec:autoencoders} and measure the bit error rate and the block error rate at the receiver as outlined in Section~\ref{sec:performance}.  

\vspace*{-1em}
\section{Autoencoders}\label{sec:autoencoders}
Deep autoencoders are a ``self-supervised'' learning technique that uses deep neural networks to efficiently learn how to compress and encode data.  Then they learn to reconstruct the data back from the compressed encoded representation to a representation that is close to the original input (hence self-supervised).  They are a lossy compression technique since the reconstructed data is not equal to the original.  Autoencoders are comprised of three different layers: an encoder, a latent layer, and a decoder.  An autoencoder is shown in Fig.~\ref{fig:autoencoders}.  

\begin{figure}[!t]
\centering
\resizebox{0.35\textwidth}{!}{\begin{tikzpicture}[thick, scale=0.5, node distance=2em, >=latex, cnode/.style={draw=black,fill=#1,minimum width=0.3cm,circle}]
    \draw[fill=red!20, rounded corners=10pt] (1,1) rectangle (7,-11) node[left, xshift=-2em, yshift=-1em] {Encoder};
    \draw[fill=green!20, rounded corners=10pt] (7.4,1) rectangle (10.5,-11) node[left, xshift=-0.5em, yshift=-1em] {Latent};
    \draw[fill=blue!20, rounded corners=10pt] (10.9,1) rectangle (16.5,-11) node[left, xshift=-2em, yshift=-1em] {Decoder};
    
    \foreach \x in {1,...,4}
    {
      \pgfmathparse{\x== 3 ? "\vdots" : "\hat h_\x"}
             \pgfmathparse{\x== 4? "\hat h_{N}" : "\pgfmathresult"}
        \node[cnode=gray!20,label=180:${\pgfmathresult}$] (x-\x) at (0,{-2*\x}) {};
        \pgfmathparse{\x== 3 ? "\vdots" : "\hat{\hat h}_\x"}
             \pgfmathparse{\x== 4? "\hat{\hat h}_{N}" : "\pgfmathresult"}         
         \node[cnode=gray!20,label=0:${\pgfmathresult}$] (xn-\x) at (18,{-2*\x}) {};
    }
    
    \foreach \x in {1,...,6}
    {
        \node[cnode=gray!10] (h1-\x) at (3,{-2*\x+2}) {}; 
        \node[cnode=gray!40] (hn1-\x) at (15,{-2*\x+2}) {};
        
    }

    \foreach \x in {1,...,3}
    {    
        \node[cnode=gray!10] (h2-\x) at (6,{-2*\x-1}) {};
        \node[cnode=gray!40] (hn2-\x) at (12,{-2*\x-1}) {};
    }

    \foreach \x in {1,2}
    {    
        \node[cnode=gray] (h3-\x) at (9,{-2*\x-2}) {};
    }
    
    \foreach \x in {1,...,4}
    {   
        \foreach \y in {1,...,6}
        {  
            \draw (x-\x) -- (h1-\y); 
            \draw (xn-\x) -- (hn1-\y); 
            
        }
    }
    \foreach \x in {1,...,6}
    {   
        \foreach \y in {1,...,3}
        {  
            \draw (h1-\x) -- (h2-\y);
            \draw (hn1-\x) -- (hn2-\y);
        }
    }

    \foreach \x in {1,...,3}
    {   
        \foreach \y in {1,2}
        {  
            \draw (h2-\x) -- (h3-\y);
            \draw (hn2-\x) -- (h3-\y);
        }
    }
    
\end{tikzpicture}}%
\caption{An autoencoder has three different components: an encoder, a latent layer, and a decoder.}
\label{fig:autoencoders}
\end{figure}

\textbf{Autoencoders for channel (de-)compression:} Let us define a channel compression ratio $\kappa\colon 0 < \kappa < 1$ for the exploitation of the autoencoder.  The encoder reduces the dimension of the channel to a latent dimension and transmits it before the decoder attempts to restore channel from its latent dimension.  \review{The dimension of the latent layer is agreed upon both the UE and gNB: a smaller latent dimension enables transmission of less bits}.  Also, we write $\mathbf{\hat h}_v\coloneqq \text{vec}(\mathbf{\hat H}_v)$ as a column vector of this CSI matrix.  This vector has the dimension of $N_\text{sc}N_rN_t$.  The UE compresses the channel using the encoder to a dimension $D\coloneqq N_rN_t\lceil(1 - \kappa)N_\text{sc}\rceil$ while the BS decompresses it and reconstructs the CSI using the decoder.

\textbf{Complex-aware neural networks:}  Neural networks are non-linear operators due to the non-linear activation functions;  therefore they cannot be applied on the real and imaginary part of a complex number independently.  Thus, we propose $U\colon \mathbb{C}^\mu\rightarrow\mathbb{R}^{2\mu}$ as a transformation. This allows us to write $\mathbf{x} \mapsto [\text{Re}(\mathbf{x}), \text{Im}(\mathbf{x})]^\top$, effectively doubling the dimension of $\mathbf{x}$ to $2N_t$ from $N_t$.  Because of this, the autoencoder reduces the dimension of the vectorized CSI to $2 N_rN_t \lceil (1 - \kappa) N_\text{sc}\rceil$ from the original $2N_\text{sc}N_rN_t$.  It should also be understood that the further the latent dimension is reduced, the more the loss information happens after reconstruction.


\textbf{Hyperparameters:} Each layer of the autoencoder is comprised of several neurons, each with a choice of rectified linear (ReLU) or sigmoid activation functions.  The hyperparameters of each layer $\boldsymbol\Theta$ comprise the weights of the neurons across the depth and width of each layer. The adaptive moments \cite{KingmaB14} (Adam) variant of gradient descent is used to optimize the cost (or loss) function, which is a complex number aware  mean square error function of both the vectorized original uncompressed estimated channel and the vectorized reconstructed (or decompressed) channel estimate, defined as:
\begin{equation}\label{eq:mse}
    \mathsf{MSE}(\mathbf{\hat h}_v, \mathbf{\hat{\hat h}}_v; \boldsymbol \Theta) \coloneqq \frac{1}{N_\text{sc}N_rN_t} \sum\nolimits_i \vert {\hat h}_v^{(i)} - \hat{\hat h}_v^{(i)} \vert ^ 2.
\end{equation}%

This optimization is done as part of the deep neural network training, which is defined by the number of training epochs and a batch size on which the gradient descent takes place.

\review{
\textbf{Training:} The training of the autoencoder is done using a large sample of channels from the users in the service area of the BS for sufficient run time.  This is necessary for the compression to be aware of any variations in the CSI.  Tuning relevant hyperparameters such as the batch size, choice of optimizer, and learning rate is necessary. 
}

\textbf{Run-time complexity:} Fixing the number of training epochs, the batch size, the depth and the width of the deep neural networks in the autoencoder, and the wireless system parameters, the run-time complexity is in $\mathcal{O}(1)$---independent of the compression ratio.%

\review{
\begin{figure}[!t]
    \centering
    \resizebox{0.45\textwidth}{!}{\begin{tikzpicture}[style=thick, >=latex]

\draw[->] (-0.2,0) -- ++(9.2,0) node[right] {$t$};

\node[yshift=0.675em, xshift=1.25em, draw, text width=3em, text centered, minimum height=1em,fill=white!80!black](learn) {Learn};

\node[right=of learn, xshift=-2.88em, draw, text width=8em, text centered, minimum height=1em](infer) {Infer};
\node[right=of infer,xshift=-2.88em, draw, text width=3em, text centered, minimum height=1em,fill=white!80!black](learn2) {Learn};
\node[right=of learn2, xshift=-2.88em, draw, text width=8em, text centered, minimum height=1em](infer2) {Infer};

\draw[->] (-0.2,-1) -- ++(9.2,0) node[right] (input-1) {$t$};

\foreach \x in {0,1,...,8} {
\pgfmathsetmacro{\col}{ifthenelse(Mod(\x, 4) == 0, "white!80!black", "transparent!0"))}
\pgfmathsetmacro{\y}{int(8-\x)}
\pgfmathsetmacro{\counter}{ifthenelse(\x < 3, "\x", ifthenelse(\y == 0, "i", "i-\y"))}

\ifthenelse{\x = 3}{
    \node [rectangle, text width=2em, text centered, text height=1em] at (0.72em+2.7*\x em, -0.675)  {$\cdots$};
  }{
    \node [rectangle, draw, fill=\col, text width=2em, text centered, text height=1em] at (0.72em+2.7*\x em, -0.675) {$h_{\counter}$};
  }
}
\end{tikzpicture}}%
    \caption{Timing diagram showing two training-inference time patterns: duty cycle (top) and staggering (bottom).}\label{fig:train_inf}
\end{figure}%
}

\begin{figure*}[!t]
    \centering
    \subfloat[Original (true, uncompressed).]{\includegraphics[width=0.3\textwidth]{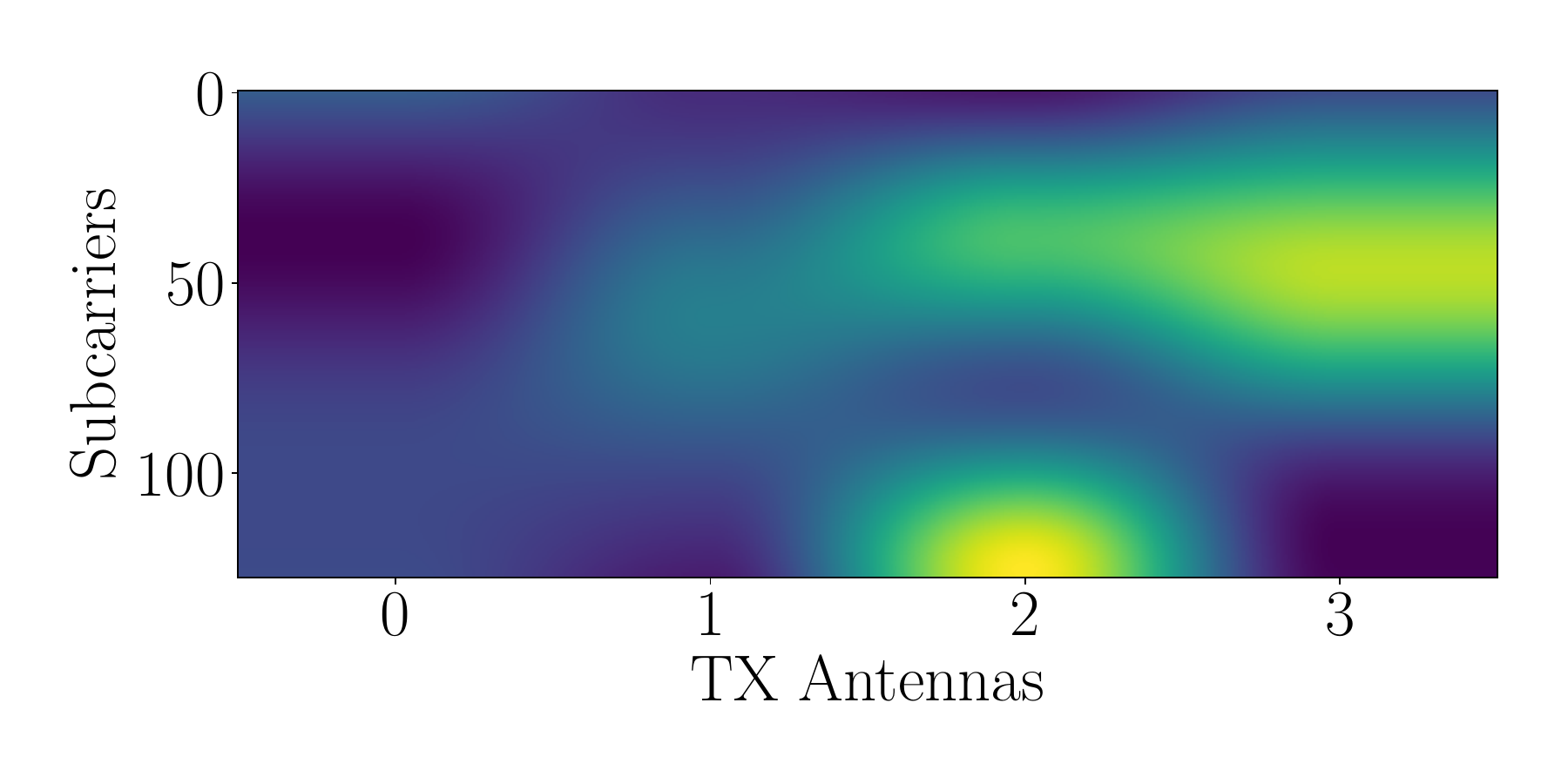}}%
    \hfil
    \subfloat[Compressed $\kappa=0.7$.]{\includegraphics[width=0.3\textwidth]{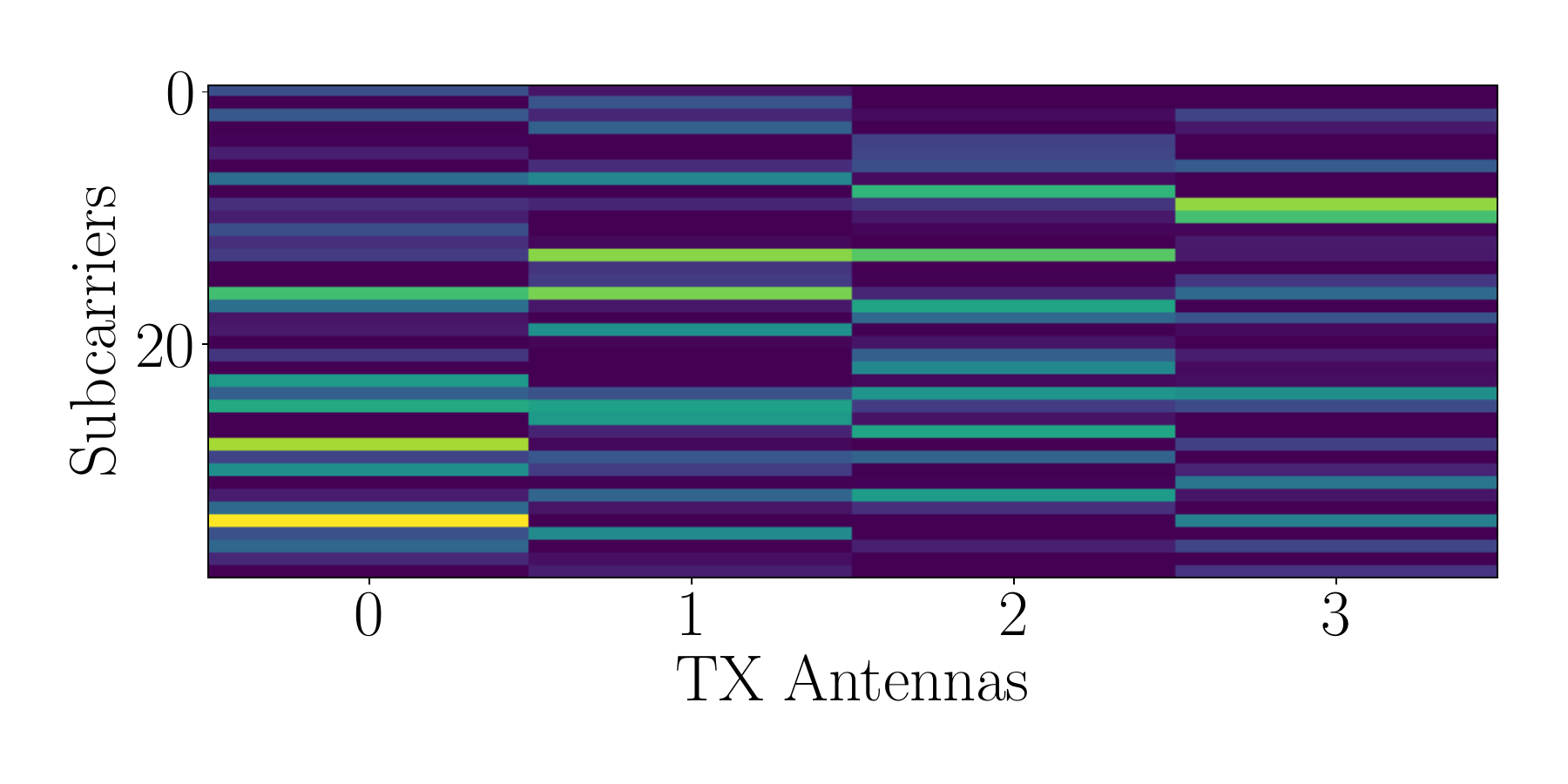}}%
    \hfil
    \subfloat[Reconstructed $\kappa=0.7$.]{\includegraphics[width=0.3\textwidth]{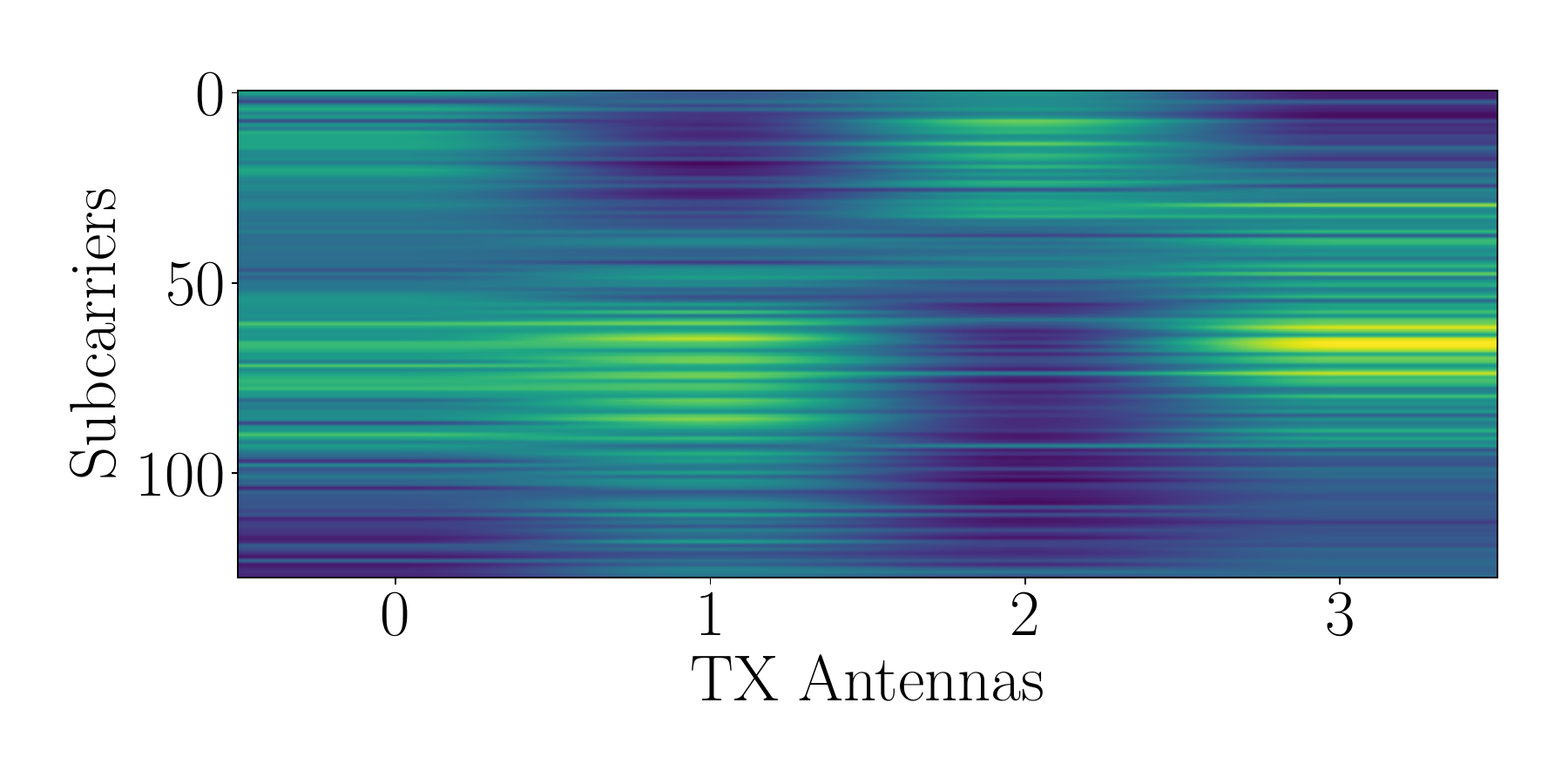}}%
    \caption{The CDL-E channel state information matrix compression rate $\mathbf{H}$ \review{of a small ($4\times 4$) uniform rectangular antenna} for the first receive antenna and a given user under different scenarios: uncompressed, compressed, and reconstructed at transmit SNR $\rho = 30$ dB.}%
    \label{fig:channels}
\end{figure*}%

\review{
\vspace*{-0.5em}
\section{Practical Considerations}\label{sec:practical}%
\textbf{Adaptivity:} Since the run-time complexity is indifferent to the compression ratio, an \textit{adaptive} algorithm that switches between different compression ratios $\mathcal{K} \coloneqq \{\kappa_1, \kappa_2, \ldots\}$ based on the computed performance measure is possible.  Here, the encoder compresses the CSI using the ``best'' compression ratio for the chosen performance measure.  This can be implemented either via a lookup table (e.g., similar to \cite{3gpp38214}) or by using a supervised deep learner functioning as a classifier. This adaptivity implements this optimization problem:%
\begin{equation}\label{eq:adaptive}
    \kappa^\star \coloneqq \underset{\kappa\in\mathcal{K}}{\arg\,\min}\;\pi(\mathbf{x}, \mathbf{\hat x}; \kappa),
\end{equation}%
where $\pi(\cdot)$ is an error measure of the received codeword $\mathbf{\hat x}$ as defined in Section~\ref{sec:performance}. Every BS transmits a \textit{sequence} of pilots (note how this differs from Section~\ref{sec:sysmodel}) each with a known compression ratio from $\mathcal{K}$ as in Fig.~\ref{fig:train_inf} during the training period.  Then, based on the performance measure computed by the UE on the subsequent radio frame transmitted from the serving BS, the serving BS constructs a dataset $\mathcal{D}$ with the reported performance measure for each served UE and the compression ratio from $\mathcal{K}$ as a supervisory signal.  Information related to the UE location, angle of departure, and received signal level can also further enhance the predictability of the deep learner.  In simple terms, $\mathcal{D} \coloneqq [\mathbf{X} \coloneqq [\rho, \pi, \ldots] \mid \mathbf{y} \coloneqq \kappa]$.  During the inference period, the deep classifier thus generates a class from $\mathcal{K}$ as a solution for \eqref{eq:adaptive} at the current radio frame.  %

\textbf{Signaling overhead:}  In order for compression to work as intended, both the UE and the BS must be made aware of the compression ratio a CSI has been processed with during the present radio frame.  Therefore, the control plane channels needs to accommodate transmitting an additional overhead.  This overhead has parts that belong to the encoder (i.e., the compressed CSI) and the bits the BS uses to find out the compression ratio used by the UE to perform the decompression.  Computationally, for a compressed CSI of dimension $D$ and a number of required bits per CSI element $b \in \mathbbm{Z}_+$, which and depends on the hardware architecture, a total of $bD$ bits is necessary to represent the compressed CSI.  Therefore, the overhead due to the adaptivity becomes $bD + \lceil \log_2 \vert\mathcal{K}\vert \rceil$ bits, implying that the overhead scales in a log complexity as a function of the number of available compression ratios.%

\textbf{Training-inference patterns:} The inference stage enables the trained autoencoder to use new CSI data it may not have been trained on and make predictions.  The inference stage uses the CSI data reported by other UEs to compute the inference loss.  For the duration of a given radio frame, Fig.~\ref{fig:train_inf} shows two possible patterns: a) duty cycle where the loss is observed over the inference time period and b) staggering where a measurement occasion parameter specifies which non-contiguous time slots are used within a radio frame to train the autoencoder and which other non-contiguous time slots are used for the inference (and the loss observation).  In both cases, if the inference loss is above a certain threshold, invalidation and retraining of the autoencoder is necessary \cite{bandswitching}.


}

\vspace*{-0.5em}
\section{Performance Measures}\label{sec:performance}%
\textbf{Bit error rate}: We define the empirical bit error rate (BER) as the ratio of bits that were received incorrectly (compared to the transmitted ones) as a result of the channel compression, fading, and noise.  Let a given codeword of the $n$-th transmission be designated by a string of bits $\mathbf{x}_n\coloneqq (x_{n, 1}, x_{n, 2}, \ldots, x_{n, W})$.  Therefore for $N_\text{trans}$ number of transmissions for a given UE:
\begin{equation}
p_b\review{(\mathbf{x}, \mathbf{\hat x}; \kappa)} \coloneqq \frac{1}{N_\text{trans}W} \sum_{n=1}^{N_\text{trans}}\sum_{i=1}^W \mathbbm{1} [x_{n, i} \neq \hat x^{(\kappa)}_{n, i}]
\end{equation}%

\textbf{Block error rate}: Further we compute the block error rate (BLER) defined as the probability of the CRC mismatch between the transmitted and received codeword based on the number of transmissions:
\begin{equation}
\begin{aligned}\label{eq:bler}
    \textsf{BLER}\review{(\mathbf{x}, \mathbf{\hat x}; \kappa)} & \coloneqq \mathbb{P}[\mathbf{c}(\mathbf{x}) \neq \mathbf{c}(\mathbf{\hat x}^{(\kappa)})] \\
    &= \frac{1}{N_\text{trans}}\sum_{n=1}^{N_\text{trans}} \mathbbm{1} [\mathbf{c}(\mathbf{x}_n) \neq \mathbf{c}(\mathbf{\hat x}_n)]
\end{aligned}
\end{equation}
The CRC is calculated based on a generator polynomial $\mathbf{c}(\cdot)$ with a finite length.  It is appended to the end of every transmitted codeword regardless of the number of streams, with zero-padding as necessary to ensure that the codewords divide $N_\text{sc}N_s$ (without a remainder).%

\vspace*{-0.5em}
\section{Simulation}\label{sec:simulation}%
\vspace*{-0.5em}
\subsection{Setup}
We simulate the system outlined in Section~\ref{sec:sysmodel} with the parameters in Table~\ref{tab:rfparameters}.  Users are represented as different random seeds in the simulation \cite{mismar2023quick}. The equalizer $\mathbf{W}$ is the minimum mean squared error (MMSE) equalizer. We construct an autoencoder with the hyperparameters shown in Table~\ref{tab:autoencoderparams} in addition to a learning rate of $10^{-4}$, an epoch count of $64$, and a batch size of $128$.  Further we simulate several compression ratios $\kappa \in \mathcal{K} \coloneqq \{0.1, 0.5, 0.7\}$  \review{for two URA sizes: small ($4\times 4$) and large ($256\times 4$).  We also simulate an adaptive compression scheme which optimizes the performance measure as in \eqref{eq:adaptive}.}  We assume full-buffer transmission and choose a \review{$9.6$-megabyte} bitmap image as the transmitted payload. The quantizer function $Q(\cdot)$ is set to truncate past the sixth decimal.  We simulate for a set of transmit SNRs, $\rho \in \{0,5,\ldots,30\}$ \review{using the duty cycle training approach} for performance insights.

\begin{table}[!t]
\centering
\setlength\doublerulesep{0.5pt}
\caption{System Parameters}
\vspace*{-0.5em}
\label{tab:rfparameters}
\small
\begin{tabular}{ lr } 
\hhline{==}
Parameter & Value \\
\hline
Channel model & CDL-C and CDL-E \cite{3gpp38900}\\
Modulation & $16$-QAM\\
\review{Transmit URA configuration (small)} & \review{$4\times 4$} \\
\review{Transmit URA configuration (large)} & \review{$256\times 4$} \\
Number of receive antennas $N_r$ & $4$ \\
Number of OFDM subcarriers $N_\text{sc}$ & $128$ \\
Number of pilot symbols $N_\text{pilot}$ & $64$ \\
CRC length & $8$ bits \\
Number of users $N_\text{UE}$ & $10$ \\
Center frequency & $2100$ MHz \\
Subcarrier spacing $\Delta f$ & $15$ kHz\\
\hhline{==}
\end{tabular}
\end{table}%

\begin{table}[!t]
\centering
\setlength\doublerulesep{0.5pt}
\caption{Autoencoder Hyperparameters}
\vspace*{-0.5em}
\label{tab:autoencoderparams}
\small
\begin{tabular}{ llr } 
\hhline{===}
Layer & Parameter & Value \\
\hline
Encoder &  Depth & $2$\\
 & Width & $(10, 10)$\\
 & Activation & (ReLU, ReLU) \\
Latent layer & Width & $2N_rN_t \lceil  (1 - \kappa)  N_\text{sc} \rceil $\\
Decoder & Depth & $2$ \\
 & Width & $(10, N_rN_tN_\text{sc})$\\ 
 & Activation & (ReLU, sigmoid) \\
\hhline{===}
\end{tabular}
\end{table}%


\begin{figure*}[!t]
    \centering
    \subfloat[Bit error rates (small, CDL-E)]{\resizebox{0.24\textwidth}{!}{
\begin{tikzpicture}

\begin{semilogyaxis}[
width=8in,
height=5in,
tick align=outside,
tick pos=left,
x grid style={white!69.0196078431373!black},
xlabel={Transmit SNR $\rho$ [dB]},
xmajorgrids,
xmin=0, xmax=30,
xtick={0,5,...,30},
xtick style={color=black},
y grid style={white!69.0196078431373!black},
ylabel={Bit Error Rate $p_b$},
ymajorgrids,
ymin=1e-3, ymax=1,
scale=0.36,
yminorgrids,
ytick style={color=black},
legend columns=2,
legend cell align={left},
legend style={at={(0.975,0.28)}, fill opacity=0.6, draw opacity=1, text opacity=1, draw=white!80!black, nodes={scale=1}},
]
\addplot [thick, black, opacity=0.7, dashed, mark=*, mark size=2, mark options={solid}] 
table {%
0 0.616699735449735
5 0.519494047619048
10 0.373462301587302
15 0.202178406084656
20 0.0776992394179894
25 0.0160176917989418
30 0.000599371693121693
};
\addlegendentry{CDL-E Baseline};

\addplot [thick, green, opaque=0.5, mark=pentagon*, mark size=4, mark options={solid}]
table {%
0 0.642642195767196
5 0.59692046957672
10 0.525231481481481
15 0.418696263227513
20 0.295477843915344
25 0.171073082010582
30 0.0767774470899471
};
\addlegendentry{CDL-E $\kappa = 0.1$};

\addplot [thick, red, opaque=0.7, mark=triangle*, mark size=2, mark options={solid}]
table {%
0 0.654980985449735
5 0.609275793650794
10 0.535358796296296
15 0.432146990740741
20 0.308920304232804
25 0.186553406084656
30 0.0946263227513228
};
\addlegendentry{CDL-E $\kappa = 0.5$};

\addplot [thick, blue, opaque=0.5, mark=diamond*, mark size=4, mark options={solid}]
table {%
0 0.62880291005291
5 0.572937334656085
10 0.500181878306878
15 0.409656084656085
20 0.307676091269841
25 0.205898644179894
30 0.114996693121693
};
\addlegendentry{CDL-E $\kappa = 0.7$};

\end{semilogyaxis}
\end{tikzpicture}}}%
    \hfil
    \subfloat[Bit error rates (small, CDL-C)]{\resizebox{0.24\textwidth}{!}{
\begin{tikzpicture}

\begin{semilogyaxis}[
width=8in,
height=5in,
tick align=outside,
tick pos=left,
x grid style={white!69.0196078431373!black},
xlabel={Transmit SNR $\rho$ [dB]},
xmajorgrids,
xmin=0, xmax=30,
xtick={0,5,...,30},
xtick style={color=black},
y grid style={white!69.0196078431373!black},
ylabel={Bit Error Rate $p_b$},
ymajorgrids,
ymin=1e-3, ymax=1,
scale=0.36,
yminorgrids,
ytick style={color=black},
legend columns=2,
legend cell align={left},
legend style={at={(0.975,0.28)}, fill opacity=0.6, draw opacity=1, text opacity=1, draw=white!80!black, nodes={scale=1}},
]

\addplot [thick, black, opacity=0.7, dashed, mark=*, mark size=2, mark options={solid}] 
table {%
0 0.636065641534391
5 0.549925595238095
10 0.417042824074074
15 0.25129794973545
20 0.11067294973545
25 0.0330026455026455
30 0.00419973544973545
};
\addlegendentry{CDL-C Baseline};

\addplot [thick, green, opaque=0.5, mark=pentagon*, mark size=4, mark options={solid}]
table {%
0 0.639914021164021
5 0.582593419312169
10 0.498842592592593
15 0.393353174603175
20 0.271147486772487
25 0.153604497354497
30 0.0648809523809524
};
\addlegendentry{CDL-C $\kappa = 0.1$};

\addplot [thick, red, opaque=0.7, mark=triangle*, mark size=2, mark options={solid}]
table {%
0 0.644146825396825
5 0.604484953703704
10 0.533928571428571
15 0.426347552910053
20 0.295581183862434
25 0.165009093915344
30 0.0726603835978836
};
\addlegendentry{CDL-C $\kappa = 0.5$};

\addplot [thick, blue, opaque=0.5, mark=diamond*, mark size=4, mark options={solid}]
table {%
0 0.646250826719577
5 0.599822255291005
10 0.523631779100529
15 0.418803736772487
20 0.292113095238095
25 0.172949735449735
30 0.0843377976190476
};
\addlegendentry{CDL-C $\kappa = 0.7$};

\end{semilogyaxis}
\end{tikzpicture}}}%
    \hfil
    \subfloat[Bit error rates (large, CDL-E)]{\resizebox{0.24\textwidth}{!}{
\begin{tikzpicture}

\begin{semilogyaxis}[
width=8in,
height=5in,
tick align=outside,
tick pos=left,
x grid style={white!69.0196078431373!black},
xlabel={Transmit SNR $\rho$ [dB]},
xmajorgrids,
xmin=0, xmax=30,
xtick={0,5,...,30},
xtick style={color=black},
y grid style={white!69.0196078431373!black},
ylabel={Bit Error Rate $p_b$},
ymajorgrids,
ymin=1e-3, ymax=1,
scale=0.36,
yminorgrids,
ytick style={color=black},
legend columns=2,
legend cell align={left},
legend style={at={(0.975,0.28)}, fill opacity=0.6, draw opacity=1, text opacity=1, draw=white!80!black, nodes={scale=1}},
]
\addplot [thick, black, opacity=0.7, dashed, mark=*, mark size=2, mark options={solid}] 
table {%
0 0.616699735449735
5 0.519494047619048
10 0.373462301587302
15 0.202178406084656
20 0.0776992394179894
25 0.0160176917989418
30 0.000599371693121693
};
\addlegendentry{CDL-E Baseline};

\addplot [thick, green, opaque=0.5, mark=pentagon*, mark size=4, mark options={solid}]
table {%
0 0.634230324074074
5 0.578563161375661
10 0.491596395502646
15 0.376884920634921
20 0.254228670634921
25 0.13776455026455
30 0.0529348544973545
};
\addlegendentry{CDL-E $\kappa = 0.1$};

\addplot [thick, red, opaque=0.7, mark=triangle*, mark size=2, mark options={solid}]
table {%
1 0.649049272486772
5 0.603786375661376
10 0.535040509259259
15 0.432047784391534
20 0.305047123015873
25 0.172949735449735
30 0.0770957341269841
};
\addlegendentry{CDL-E $\kappa = 0.5$};

\addplot [thick, blue, opaque=0.5, mark=diamond*, mark size=4, mark options={solid}]
table {%
0 0.648330026455026
5 0.607676091269841
10 0.536177248677249
15 0.433672288359788
20 0.310491071428571
25 0.19239417989418
30 0.103571428571429
};
\addlegendentry{CDL-E $\kappa = 0.7$};

\end{semilogyaxis}
\end{tikzpicture}}}%
    \hfil
    \subfloat[Bit error rates (large, CDL-C)]{\resizebox{0.24\textwidth}{!}{
\begin{tikzpicture}

\begin{semilogyaxis}[
width=8in,
height=5in,
tick align=outside,
tick pos=left,
x grid style={white!69.0196078431373!black},
xlabel={Transmit SNR $\rho$ [dB]},
xmajorgrids,
xmin=0, xmax=30,
xtick={0,5,...,30},
xtick style={color=black},
y grid style={white!69.0196078431373!black},
ylabel={Bit Error Rate $p_b$},
ymajorgrids,
ymin=1e-3, ymax=1,
scale=0.36,
yminorgrids,
ytick style={color=black},
legend columns=2,
legend cell align={left},
legend style={at={(0.975,0.28)}, fill opacity=0.6, draw opacity=1, text opacity=1, draw=white!80!black, nodes={scale=1}},
]
\addplot [thick, black, opacity=0.7, dashed, mark=*, mark size=2, mark options={solid}] 
table {%
0 0.658420138888889
5 0.595329034391534
10 0.484961970899471
15 0.319374173280423
20 0.13172123015873
25 0.0236731150793651
30 0.000830853174603175
};
\addlegendentry{CDL-C Baseline};

\addplot [thick, green, opaque=0.5, mark=pentagon*, mark size=4, mark options={solid}]
table {%
0 0.648404431216931
5 0.607188326719577
10 0.536611276455026
15 0.429799107142857
20 0.298152281746032
25 0.167435515873016
30 0.0676793981481481
};
\addlegendentry{CDL-C $\kappa = 0.1$};

\addplot [thick, red, opaque=0.7, mark=triangle*, mark size=2, mark options={solid}]
table {%
0 0.618816137566138
5 0.560321593915344
10 0.485218253968254
15 0.39239417989418
20 0.285755621693122
25 0.17250744047619
30 0.0795758928571429
};
\addlegendentry{CDL-C $\kappa = 0.5$};

\addplot [thick, blue, opaque=0.5, mark=diamond*, mark size=4, mark options={solid}]
table {%
0 0.654385747354497
5 0.614335317460317
10 0.54796626984127
15 0.448875661375661
20 0.329621362433862
25 0.210763888888889
30 0.116728670634921
};
\addlegendentry{CDL-C $\kappa = 0.7$}; 

\end{semilogyaxis}
\end{tikzpicture}}}%
    \hfil
    \caption{Bit error rates as a function of the transmit SNR $\rho$ for \review{for both the small and large URA configurations for} for both channel types and several values of $\kappa$ compared to the baseline (no compression).}
    \label{fig:errorrates_ura}
\end{figure*}%

\begin{figure}[!t]
\centering
\subfloat[Normalized run time]{\resizebox{0.24\textwidth}{!}{
\begin{tikzpicture}

\begin{axis}[
width=4in,
height=2.5in,
legend columns=2,
legend cell align={left},
legend style={fill opacity=0.8, draw opacity=1, text opacity=1, draw=white!80!black},
tick align=outside,
tick pos=left,
scale=0.5,
x grid style={white!69.0196078431373!black},
xlabel={Compression ratio ($\kappa$)},
xmajorgrids,
xmin=0, xmax=1,
xtick={0.1,0.3,...,1},
xminorgrids,
xtick style={color=black},
y grid style={white!69.0196078431373!black},
ylabel={Normalized run time},
ymajorgrids,
ymin=0.4, ymax=1.1,
yminorgrids,
ytick style={color=black}
]

\addplot [thick, blue, mark=*]
table {%
0.1 1
0.2 1
0.3 1
0.4	1
0.5	1.009745922
0.6	0.992738892
0.7	0.994695858
0.8	0.986611709
0.9	1.010386651
};
\end{axis}

\end{tikzpicture}}}%
\hfil
\subfloat[Training and validation losses]{\resizebox{0.24\textwidth}{!}{
\begin{tikzpicture}

\begin{axis}[
width=4in,
height=2.5in,
legend columns=1,
legend cell align={left},
legend style={fill opacity=0.8, draw opacity=1, font=\Large, text opacity=1, draw=white!80!black, nodes={scale=0.52, transform shape}},
tick align=outside,
tick pos=left,
scale=0.5,
x grid style={white!69.0196078431373!black},
xlabel={Epoch},
xmajorgrids,
xmin=0, xmax=64,
xtick={0,20,...,60},
xminorgrids,
xtick style={color=black},
y grid style={white!69.0196078431373!black},
ylabel={Loss $\mathsf{MSE}(\mathbf{h}_v, \mathbf{\hat h}_v)$},
ymajorgrids,
ymin=0, ymax=2.05,
yminorgrids,
ytick style={color=black}
]

\addplot [thick, blue]
table {%
0 1.99206280708313
1 1.98648238182068
2 1.97227919101715
3 1.94223475456238
4 1.89186227321625
5 1.82110011577606
6 1.73904991149902
7 1.66630077362061
8 1.59347605705261
9 1.49433565139771
10 1.39110291004181
11 1.29915654659271
12 1.20424151420593
13 1.10053813457489
14 1.00430154800415
15 0.920831203460693
16 0.831767320632935
17 0.745411098003387
18 0.676032364368439
19 0.60981696844101
20 0.548078656196594
21 0.501598715782166
22 0.452819108963013
23 0.409418940544128
24 0.377509713172913
25 0.342771410942078
26 0.317313969135284
27 0.288346409797668
28 0.264457851648331
29 0.241535782814026
30 0.222786605358124
31 0.203336223959923
32 0.188079029321671
33 0.169988945126534
34 0.1587965041399
35 0.145822986960411
36 0.134322792291641
37 0.127931267023087
38 0.121078871190548
39 0.111495815217495
40 0.101460635662079
41 0.09354467689991
42 0.0891664102673531
43 0.0921992063522339
44 0.0964129790663719
45 0.0785380601882935
46 0.0710952058434486
47 0.0792646482586861
48 0.0750531703233719
49 0.0594080612063408
50 0.0582654736936092
51 0.069433867931366
52 0.0580417886376381
53 0.045851182192564
54 0.0459109619259834
55 0.0613216310739517
56 0.0514713674783707
57 0.0386252366006374
58 0.0373490527272224
59 0.055045872926712
60 0.0464088283479214
61 0.0323357917368412
62 0.0278736613690853
63 0.0377214774489403
};
\addlegendentry{CDL-E training loss}

\addplot [thick, red]
table {%
0 1.98648238182068
1 1.97227919101715
2 1.94223475456238
3 1.89186227321625
4 1.82110011577606
5 1.73904991149902
6 1.66630077362061
7 1.59347605705261
8 1.49433565139771
9 1.39110291004181
10 1.29915654659271
11 1.20424151420593
12 1.10053813457489
13 1.00430154800415
14 0.920831203460693
15 0.831767320632935
16 0.745411098003387
17 0.676032364368439
18 0.60981696844101
19 0.548078656196594
20 0.501598715782166
21 0.452819108963013
22 0.409418940544128
23 0.377509713172913
24 0.342771410942078
25 0.317313969135284
26 0.288346409797668
27 0.264457851648331
28 0.241535782814026
29 0.222786605358124
30 0.203336223959923
31 0.188079029321671
32 0.169988945126534
33 0.1587965041399
34 0.145822986960411
35 0.134322792291641
36 0.127931267023087
37 0.121078871190548
38 0.111495815217495
39 0.101460635662079
40 0.09354467689991
41 0.0891664102673531
42 0.0921992063522339
43 0.0964129790663719
44 0.0785380601882935
45 0.0710952058434486
46 0.0792646482586861
47 0.0750531703233719
48 0.0594080612063408
49 0.0582654736936092
50 0.069433867931366
51 0.0580417886376381
52 0.045851182192564
53 0.0459109619259834
54 0.0613216310739517
55 0.0514713674783707
56 0.0386252366006374
57 0.0373490527272224
58 0.055045872926712
59 0.0464088283479214
60 0.0323357917368412
61 0.0278736613690853
62 0.0377214774489403
63 0.0591495931148529
};
\addlegendentry{CDL-E validation loss}


\addlegendentry{CDL-C validation loss}
\addplot [thick, blue, dashed]
table {%
0 2.05694842338562
1 2.04882907867432
2 2.02468180656433
3 1.97682058811188
4 1.90397000312805
5 1.82236504554749
6 1.75472748279572
7 1.65813553333282
8 1.5519654750824
9 1.4548351764679
10 1.34659337997437
11 1.22968935966492
12 1.12623584270477
13 1.02433133125305
14 0.913727283477783
15 0.818497776985168
16 0.732452750205994
17 0.649490714073181
18 0.584567964076996
19 0.518800437450409
20 0.464251399040222
21 0.419848501682281
22 0.375164836645126
23 0.340866327285767
24 0.301957279443741
25 0.273928165435791
26 0.246564149856567
27 0.224504098296165
28 0.207415834069252
29 0.187952503561974
30 0.174215137958527
31 0.164516896009445
32 0.152413979172707
33 0.139366641640663
34 0.130108624696732
35 0.125075370073318
36 0.122992172837257
37 0.118060030043125
38 0.10233897715807
39 0.092746838927269
40 0.0922172516584396
41 0.0976141169667244
42 0.0891396030783653
43 0.0724711343646049
44 0.0679120048880577
45 0.0776986107230186
46 0.0805119425058365
47 0.0599847622215748
48 0.0527041591703892
49 0.0630273371934891
50 0.0759218260645866
51 0.0473081208765507
52 0.0453408807516098
53 0.0718110352754593
54 0.047916866838932
55 0.0360339283943176
56 0.0365939661860466
57 0.0656976252794266
58 0.0417530722916126
59 0.0304159950464964
60 0.0285312421619892
61 0.05568927526474
62 0.0483761839568615
63 0.0291331652551889
};
\addlegendentry{CDL-C training loss}

\addplot [thick, red, dashed]
table {%
0 2.04882907867432
1 2.02468180656433
2 1.97682058811188
3 1.90397000312805
4 1.82236504554749
5 1.75472748279572
6 1.65813553333282
7 1.5519654750824
8 1.4548351764679
9 1.34659337997437
10 1.22968935966492
11 1.12623584270477
12 1.02433133125305
13 0.913727283477783
14 0.818497776985168
15 0.732452750205994
16 0.649490714073181
17 0.584567964076996
18 0.518800437450409
19 0.464251399040222
20 0.419848501682281
21 0.375164836645126
22 0.340866327285767
23 0.301957279443741
24 0.273928165435791
25 0.246564149856567
26 0.224504098296165
27 0.207415834069252
28 0.187952503561974
29 0.174215137958527
30 0.164516896009445
31 0.152413979172707
32 0.139366641640663
33 0.130108624696732
34 0.125075370073318
35 0.122992172837257
36 0.118060030043125
37 0.10233897715807
38 0.092746838927269
39 0.0922172516584396
40 0.0976141169667244
41 0.0891396030783653
42 0.0724711343646049
43 0.0679120048880577
44 0.0776986107230186
45 0.0805119425058365
46 0.0599847622215748
47 0.0527041591703892
48 0.0630273371934891
49 0.0759218260645866
50 0.0473081208765507
51 0.0453408807516098
52 0.0718110352754593
53 0.047916866838932
54 0.0360339283943176
55 0.0365939661860466
56 0.0656976252794266
57 0.0417530722916126
58 0.0304159950464964
59 0.0285312421619892
60 0.05568927526474
61 0.0483761839568615
62 0.0291331652551889
63 0.0247767381370068
};
\end{axis}

\end{tikzpicture}}}%
\caption{Autoencoder normalized run time vs. $\kappa$ (left) and losses vs.\ the number of epochs for a \review{large URA configuration} and static $\kappa = 0.4$ (right).}
\label{fig:dnn_perf}
\end{figure}%

\review{
\begin{figure}[!t]
\centering
\resizebox{0.3\textwidth}{!}{
\begin{tikzpicture}

\begin{semilogyaxis}[
width=8in,
height=5in,
tick align=outside,
tick pos=left,
x grid style={white!69.0196078431373!black},
xlabel={Transmit SNR $\rho$ [dB]},
xmajorgrids,
xmin=0, xmax=30,
xtick={0,5,...,30},
xtick style={color=black},
y grid style={white!69.0196078431373!black},
ylabel={Block Error Rate $\textsf{BLER}$},
ymajorgrids,
ymin=1e-5, ymax=1,
scale=0.4,
yminorgrids,
ytick style={color=black},
legend cell align={left},
legend style={at={(0.98,0.98)}, fill opacity=0.6, draw opacity=1, text opacity=1, draw=white!80!black, nodes={scale=1}},
]
\addplot [thick, black, opacity=0.7, dashed, forget plot]
table {%
0 0.1
5 0.1
10 0.1
15 0.1
20 0.1
25 0.1
30 0.1
};

\addplot [thick, black, mark=*, mark size=2, mark options={solid}] 
table {%
0 0.464444444444444
5 0.303492063492063
10 0.0514814814814815
15 5.29100529100529e-05
20 0
25 0
30 0
};
\addlegendentry{No compression};

\addplot [thick, blue, mark=square*, mark size=2, mark options={solid}] 
table {%
0 0.541322751322751
5 0.450687830687831
10 0.09892804232804233
15 0.016613756613757
20 4.4645502645502e-4
25 0
30 0
};
\addlegendentry{Static $\kappa = 0.5$};

\addplot [thick, red, opaque=0.5, mark=o, mark size=2, mark options={solid}]
table {%
10 0.07398941798
15 0.01114814815
20 2.11005291005e-5
};
\addlegendentry{Adaptive};

\end{semilogyaxis}
\end{tikzpicture}}%
\caption{\review{BLER vs $\rho$ for the adaptive compression scheme in a large URA configuration (CDL-E channel type).}}
\label{fig:adaptive_perf}
\end{figure}%
}

\vspace*{-1em}
\subsection{Discussion}
\review{\textbf{Static compression scheme:}} We start with Fig.~\ref{fig:channels} which visualizes three variants of the CSI for the CDL-E channel at the first receive antenna and \review{a static compression ratio of} $\kappa = 0.7$: original uncompressed, compressed, and reconstructed (or decompressed).   The size of data representing the subcarriers has reduced to a size equal to the latent layer dimension $\lceil(1 - \kappa)N_\text{sc}\rceil = 13$ from $N_\text{sc} = 128$.  While the reconstructed channel does not perfectly resemble the original one---since autoencoder compression is lossy---the high degree of correlation and sparsity in the CDL-E channel due to LOS propagation, provides the compression adequate redundant information for a useful reconstruction.

A question now arises:  How far better can this static compression go while maintaining its purpose?  In other words, what BER performance can be obtained for a given compression ratio $\kappa$ and transmit SNR $\rho$ \review{for various channel types and URA configurations}?  To answer this question we study Fig.~\ref{fig:errorrates_ura}.  Here, we observe that the performance of the system (as measured by the BER) degrades as $\kappa$ increases, which is an expected behavior due to the inability to perfectly equalize the channel after the lossy compression.

\review{\textbf{Adaptive compression scheme:}} We observe that at low $\rho$ values ($\rho \le 5$ dB), the received BER performance is bad even with low compression ratios.  Similarly, in the moderate regime, the uncompressed and lightly compressed channels show better performance, with the BER reducing significantly.  However, at the high $\rho$ regime ($\rho \ge 20$ dB), the BER for all compression ratios decreases.  The key insight is that as the compression ratio increases, the BER worsens due to the loss of CSI that the autoencoder fails to reconstruct with higher compression.  The adaptive compression scheme finds the ``right balance'' between compression and performance based on \eqref{eq:adaptive}.  For a more realistic comparison, we use BLER as our performance measure.  For this purpose, we use the generator polynomial $\mathbf{c}(\mathbf{x}) = x^6 + x^4 + x + 1$.  With an override of the notation in \eqref{eq:adaptive} and \eqref{eq:bler}, we thus find optimal compression for a given transmission condition $\rho$ and write
\begin{equation}
\begin{aligned}
    \kappa^\star \coloneqq  \underset{\kappa\in \mathcal{K}}{\arg\,\min} & \qquad  B\coloneqq \textsf{BLER}(\mathbf{x}, \mathbf{\hat x}; \rho, \kappa) \\
    \text{s.t.} & \qquad B \le B_\text{max}
\end{aligned}
\end{equation}
which is intractable due to the absence of a closed form of BLER \eqref{eq:bler} as a function of the compression rate $\kappa$ and SNR $\rho$.  \review{Thus, the dataset $\mathcal{D}$ constructed by the serving BS---as motiviated in Section~\ref{sec:practical}---is useful since it 1) tracks the various radio measurements and performance for the UEs it serves and 2) can train a machine learning model}.  Another feature in $\mathcal{D}$ is introduced as a flag whether the BLER has exceeded target BLER $B_\text{max} = 0.1$ per industry standards \cite{3gpp38133}.

\review{We show in} Fig.~\ref{fig:dnn_perf} that the \textit{normalized} run time of the autoencoder is approximately constant irrelative to the compression ratio.  \review{We further show} the training and validation losses of the deep autoencoder as a function of the number of training epochs for both channels.  Given that both decrease monotonically, we conclude that the autoencoder is not overfitting and thus its results can be generalized within the values assumed by the CSI. \review{This motivates the adaptive compression scheme, which we show next.}

\review{Finally, in Fig.~\ref{fig:adaptive_perf} the adaptive compression scheme chooses from various compression ratios based on the transmit SNR and conditioned upon BLER not exceeding $B_\text{max}$.  The algorithm outperforms both the scheme with no compression and a static scheme with $\kappa = 0.5$.  However, as  $\rho$ increases, the performance of the adaptive compression approaches that of the uncompressed CSI due two reasons: 1) the reconstruction of the CSI resembling the original uncompressed CSI (i.e., minimized reconstruction loss) and 2) the asymptotically efficient  equalization at high SNRs owed to MMSE.
}

\section{Conclusion}\label{sec:conclusion}
In this paper, we demonstrated the usefulness of autoencoders---as an element of a neural receiver in 6G---to compress the channel state information exploiting redundant information due to correlation.  We showed that compression is useful in high SNR regimes, which correspond to a dominant line of sight or users at the cell center.  \review{We also showed practical considerations helpful in creating an adaptive CSI compression technique.}  This procedure is a step closer towards an AI-enabled \review{massive MIMO} air interface in 6G.

\bibliographystyle{IEEEtran}
\bibliography{main.bib}

\end{document}